\documentclass[pra,twocolumn,showpacs,superscriptaddress,amsmath,amssymb]{revtex4}
\usepackage{mathrsfs}
\usepackage{graphicx}
\usepackage{dcolumn}
\usepackage{bm}

\begin{document}

\title{Physical model of continuous two-qubit parity measurement in a cavity-QED network}

\author{Joseph Kerckhoff}
\affiliation{Edward L.\ Ginzton Laboratory, Stanford University, Stanford CA 94305, USA}

\author{Luc Bouten}
\affiliation{Physical Measurement and Control 266-33, California Institute of Technology, Pasadena, CA 91125, USA}

\author{Andrew Silberfarb}
\affiliation{Edward L.\ Ginzton Laboratory, Stanford University, Stanford CA 94305, USA}

\author{Hideo Mabuchi}
\affiliation{Edward L.\ Ginzton Laboratory, Stanford University, Stanford CA 94305, USA}

\date{\today}

\begin{abstract}
We propose and analyze a physical implementation of two-qubit parity measurements as required for continuous error correction, assuming a setup in which the individual qubits are strongly coupled to separate optical cavities. A single optical probe beam scatters sequentially from the two cavities and the continuous parity measurement is realized via fixed quadrature homodyne photo-detection. We present models based on quantum stochastic differential equations (QSDE's) for both an ideal continuous parity measurement and our proposed cavity quantum electrodynamics (cavity QED) implementation; a recent adiabatic elimination theorem for QSDE's is used to assert strong convergence of the latter to the former in an appropriate limit of physical parameters. Performance of the cavity QED scheme is studied via numerical simulation with experimentally realistic parameters.
\end{abstract}


\maketitle

\noindent It is now well established~\cite{Pres98} that error correction/avoidance protocols and fault-tolerant architectures are essential for any practical implementation of quantum information processing. While most theoretical research in these areas utilizes discrete time map-based models for quantum dynamics and decoherence, which are perhaps more familiar to computer scientists and information theorists, there has been growing interest in transferring key ideas~\cite{Gott97} to the domain of continuous time differential-equation-based models, which are more common in the context of {\it ab initio} physical modeling. In this paper we contribute to a line of research, initiated by Ahn and co-workers~\cite{Ahn02,Ahn03,Ahn04,Saro04} and broadened by other research groups~\cite{VanH05,Ores07,Chas08}, which focuses on continuous quantum error correction via stabilizer coding and continuous syndrome measurement. This approach is attractive for design and analysis because it fundamentally connects the goal of quantum decoherence suppression with formal optimization methods of classical control theory~\cite{VanH05}. It also has a significant potential implementation advantage over standard discrete-time formulations in that continuous tracking of errors may be realized without the need for executing cumbersome readout circuits, but this of course relies on the assumption that continuous non-demolition syndrome measurement can be realized in an experimentally favorable way. In what follows we describe a straightforward implementation of continuous two-qubit parity measurement (sufficient for syndrome measurement of the quantum bit-flip code) in the context of cavity quantum electrodynamics (cavity QED), and analyze the performance of our scheme both for fixed realistic parameters (via numerical simulation) and in an ideal limit of parameter values (via adiabatic elimination). Our scheme utilizes a simple coherent-state optical probe in place of the usual ancillary qubits, and exploits Hamiltonian qubit-cavity couplings in place of clocked quantum logic gates for the syndrome readout. The strength of the syndrome measurement can nevertheless be modulated easily (or even turned off entirely) by adjustment of the power of the optical probe beam.

\begin{figure}
\begin{center}
\includegraphics[width=2.5in]{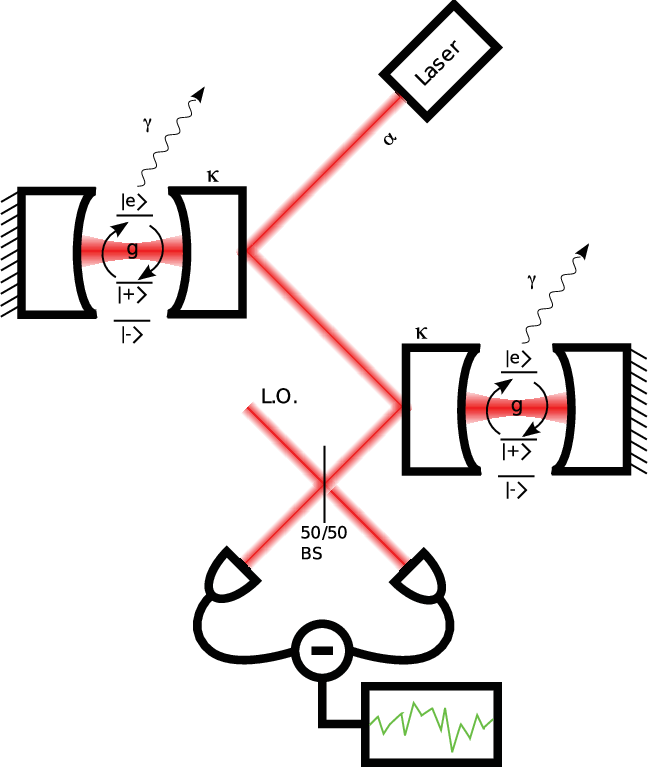}
\caption[figure1]{\label{fig:schemat} Schematic depiction of two cavities driven sequentially by a resonant laser beam. A three level atom is trapped inside each cavity, and identical atom-cavity dynamics apply in each. After probing both cavities the laser light is directed to a homodyne receiver.}
\end{center}\vspace{-0.3in}
\end{figure}

The basic setup of our proposed implementation is shown in Fig.~\ref{fig:schemat}: two optical cavities, each containing a single three-level `atom' (potentially a gas-phase alkali atom, nitrogen vacancy center in diamond, {\it etc.}), are interrogated sequentially by a coherent optical probe with amplitude $\alpha$ (similar arrangements have previously been considered in the context of quantum information science~\cite{Yama06}). A qubit is encoded in the ground states $\vert -\rangle$ and $\vert +\rangle$ of the intracavity atom; an optical transition between $\vert +\rangle$ and the excited state $\vert e\rangle$ is coupled strongly to a quantized cavity mode with vacuum Rabi frequency $g$. For simplicity we assume atomic selection rules such that $\vert e\rangle$ decays only to $\vert +\rangle$, with excited state decay rate $\gamma^2$, and single-sided cavities with photon decay rate $\kappa^2$. A homodyne receiver is used to measure the phase quadrature amplitude of the laser beam after it probes the two cavities. The basic intuition behind our scheme is that the coherent probe acquires a phase shift of either $\pi$ or $0$ radians upon reflection from each cavity, depending on whether the intracavity atom is in the coupled or uncoupled qubit state~\cite{Duan04}. After reflecting from both cavities the probe carries an overall phase shift of $\pi$ radians if the parity of the two qubits is odd, and $0$ or $2\pi$ radians if the parity is even. As the latter two conditions are indistinguishable, the homodyne measurement effectively implements a parity measurement.

We model our system quantitatively using quantum stochastic differential equations (QSDE's) and quantum filtering (quantum trajectory) theory~\cite{Bout07}. The time evolution of an observable $S$ of the atoms and fields is given by $j_t(S) = U_t^\dag SU_t$, where $U_t = U_0 + \int_0^tdU_s$ and
\begin{eqnarray}
  dU_t&=&\Bigg\{
  \left(\kappa b_1+\kappa b_2 +\alpha\right)dA^{\dag}_t
  -\left(\kappa b_1+\kappa b_2 +\alpha\right)^\dag dA_t\nonumber\\
  &&-\frac{\left(\kappa b_1 +\kappa b_2+\alpha\right)^\dag
  \left(\kappa b_1+\kappa b_2 +\alpha\right)}{2}dt\nonumber\\
  &&+\sum_{i=1}^2( \gamma\sigma^{(i)} dB^{(i)\dag}_{t}
  -\gamma\sigma^{(i)\dag}dB^{(i)}_{t} - \frac{\gamma^2\sigma^{(i)\dag}\sigma^{(i)}}{2}dt)\nonumber\\
  &&+ \frac{\kappa^2}{2}\left(b_1^{\dag} b_2 - b_2^{\dag} b_1\right)dt\ +
  g\sum_{i=1}^2\Big(\sigma^{(i)\dag}b_i - \sigma^{(i)}b_i^{\dag}\Big)dt\nonumber\\
  &&+ \left(\frac{\bar\alpha\kappa}{2} \left(b_1+ b_2\right)
  -\frac{\alpha\kappa}{2} \left(b_1+ b_2\right)^\dag\right)dt\Bigg\}U_t,\label{eq qsde}
\end{eqnarray}
with $U_0=I$. This propagator acts on $\mathfrak{h}^{\otimes 2}\otimes\mathcal{F}^{\otimes3}$, where $\mathfrak{h}=\mathbb{C}^3\otimes\ell^2(\mathbb{N})$ is the Hilbert space of a three-level atom and quantized cavity mode, and $\mathcal{F}$ is the bosonic Fock space of the probe and of the radiation modes for atomic spontaneous emission. $b_i$ and $\sigma_i$ act as the $i^{\text{th}}$ cavity mode annihilation operator and $|+\rangle\langle e|$ atomic lowering operator, respectively, and as the identity on the remaining spaces. Operators on the bosonic Fock spaces are represented as quantum noises $A_t$ and $B_t^{(i)}$ for the probe channel and $i=\{1,2\}$ radiation modes. These annihilation processes are related to the more familiar Bose fields by $A_t = \int_0^t a_sds$, $B_t^{(i)} = \int_0^t b_s^{(i)}ds$.

Eq.~\eqref{eq qsde} describes the evolution of the cavity QED system at its fundamental level, {\it i.e.}, at the level of abstraction often used in experimental considerations. Many of the more intrinsically scalable, solid state cavity QED implementations have additional dynamics not in \eqref{eq qsde}, but which could easily be incorporated via additional dephasing terms. Although complete, \eqref{eq qsde} is unwieldy with operator coefficients that couple together an infinite number of dimensions. In the strong coupling limit ($\sqrt{g}\gg\kappa,\alpha\gg\gamma$) the effective dimensionality of the system is limited to certain slow degrees of freedom and the fast dynamics can be adiabatically eliminated from the description. This is accomplished in our formalism using the theorem of singular perturbations on QSDE's~\cite{BoSi08,BoVHSi08}, from which it can be shown that the physical dynamics~\eqref{eq qsde} is approximated (with $\Pi_{12}\equiv Z_1Z_2$, $\Lambda_t = \int_0^ta_s^\dag a_sds$ and $\bar{U}_0 = Q\otimes I_{\mathcal{F}^{\otimes3}}$) by
\begin{eqnarray}\label{eq approximate qsde}
d\bar{U}_t&=&\bigl\{(\Pi_{12} - Q)d\Lambda_t + \alpha \Pi_{12}dA_t^\dag-\bar{\alpha}QdA_t
\nonumber\\
&&-\frac{|\alpha|^2}{2}Qdt\bigr\}\bar{U}_t,
\end{eqnarray}
in the sense that $\forall \psi\in \mathcal{Q}\otimes\mathcal{F}^{\otimes3}$,
\begin{equation}\label{eq:convergeU2}
 \lim_{\substack{\alpha,\kappa \to \infty\\ \frac{\alpha}{\kappa} = \mbox{const.}}} \lim_{g\to\infty}
  \left\|\left(
  U_t - \bar{U}_t\right)\psi\right\| = 0.
  \end{equation}
Here $Q$ projects onto $\mathcal{Q}$, $\mathcal{Q} = \mathcal{Q}_0^{\otimes2}$, $\mathcal{Q}_0=\text{span}\{|u\rangle,|d\rangle\}$, $|u\rangle\equiv|+\rangle\otimes|0\rangle$, $|d\rangle\equiv|-\rangle\otimes|\frac{\pm2\alpha}{\kappa}\rangle$ where the cavity mode is here represented in a coherent amplitude basis, and $Z_i$ acts as $|u\rangle\langle u|-|d\rangle\langle d|$ on the $i^{\text{th}}$ cavity system. Thus, the idealized dynamics represented by Eq.~\eqref{eq approximate qsde} live on a two-qubit Hilbert space coupled only to the probe. Note also that the qubit operators now appear as non-trivial \emph{three body operators} simultaneously coupling both qubits with probe excitations (such as $Z_1Z_2d\Lambda_t$). We have derived such effective, multi-body, non-local, qubit parity interactions as a very good approximation of the fundamental picture \eqref{eq qsde} in the strong coupling limit.

Given a QSDE model, the conditional evolution equations induced by continuous measurement can be derived straightforwardly~\cite{Bout07}. The resulting quantum filtering equation, also known as a Stochastic Schr\"odinger or Stochastic Master Equation, can be used to propagate a conditional quantum state that represents a sufficient statistic for the observer's best estimates of observables $j_t(X)$ on the basis of continuous measurement records $Y_t^{(i)}$. If the conditional state remains pure for all times we may represent it by a vector $v_t\in \mathfrak{h}^{\otimes2}$.  As complete purity of the conditional state can only be achieved if all output channels corresponding to Markovian environmental couplings are monitored with perfect efficiency, this is not a realistic assumption for laboratory implementation. However the equations so derived are very useful for simulation and analysis; analogous equations for imperfect/incomplete observation with a mixed conditional state (density operator) are easily derived as required for more practical purposes.

\begin{figure}[tb]
\begin{center}
\includegraphics[scale = .35]{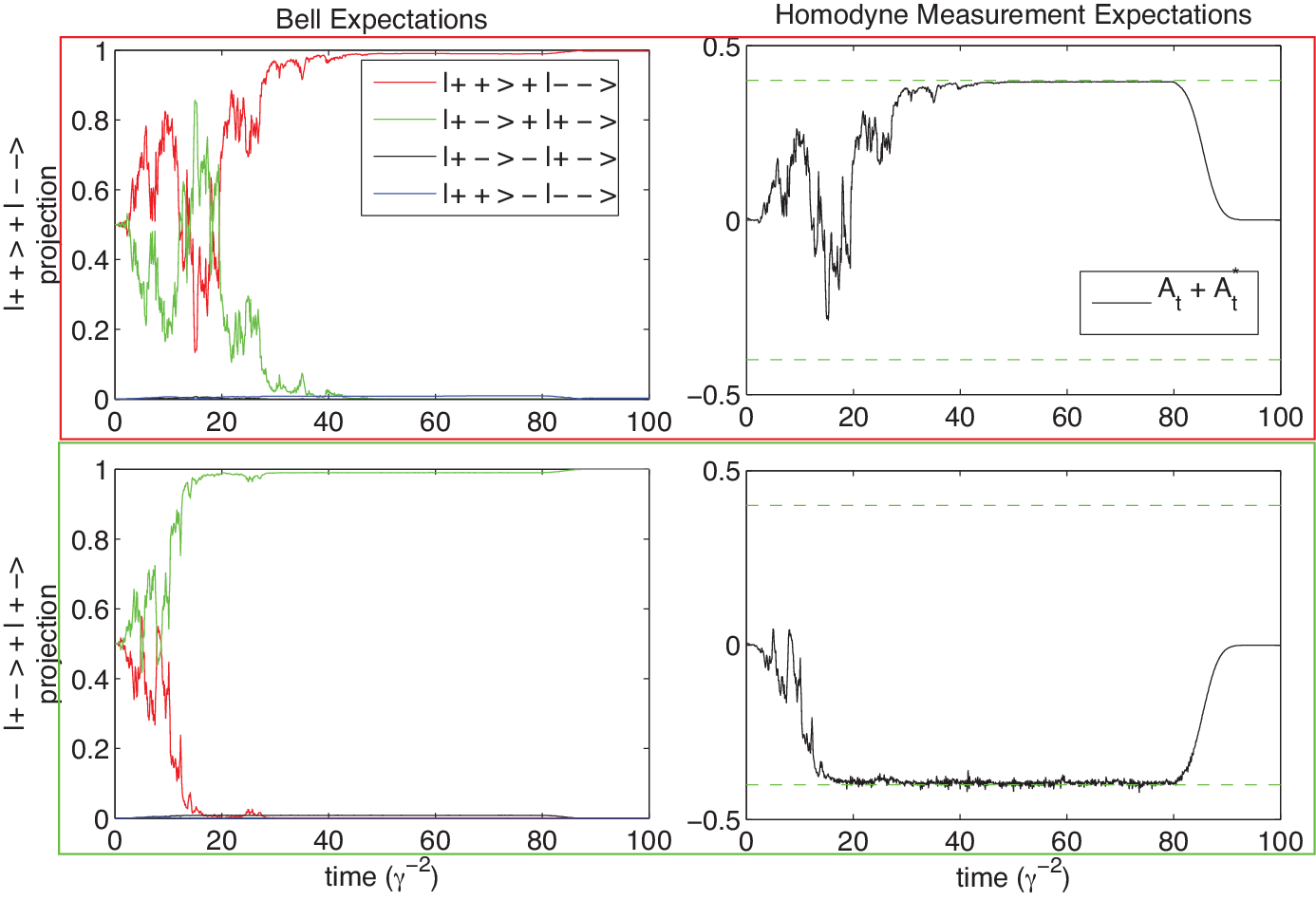}
\caption{\label{Fig 2Cavity project}
Two independent simulations of atomic Bell-state preparation in a realistic system (see text). The top pair of graphs show a simulation in which the system happened to reduce to even parity; the bottom pair shows an odd-parity example.}
\end{center}
\vspace{-0.1in}
\end{figure}

If we assume homodyne detection of the quadrature amplitude observable $Y_t^{(0)}= j_t(A_t + A_t^\dag)$ and photon counting of the radiation modes for atomic spontaneous emission $Y_t^{(k)} = j_t(\Lambda_t^{(kk)})\ k=\{1,2\}$, the `physical' pure-state propagator (for the information state $v_t$) derived from~\eqref{eq qsde} is
\begin{eqnarray}
dv_t&=&\Bigg\{(\kappa b_1 + \kappa b_2 + \alpha)dY_t^{(0)} + g\sum_{i=1}^2\Big(\sigma^{(i)\dag}b_i - \sigma^{(i)}b_i^{\dag}\Big)dt\nonumber\\
&&-(b_1^\dag b_1 + b_2^\dag b_2 )\frac{\kappa^2}{2} dt
-b_2^\dag b_1\kappa^2 dt-\alpha\kappa(b_1+b_2)^\dag dt\nonumber\\
&&+\sum_{i=1}^2\left((\gamma\sigma^{(i)}-I)dY_t^{(i)} - \frac{\gamma^2\sigma^{(i)\dag}\sigma^{(i)}}{2}dt\right)\Bigg\}v_t.\label{eq pure2}
\end{eqnarray}
For simulation purposes Eq.~\eqref{eq pure2} can be used to generate realistic homodyne measurement records $dY_t^{(0)}$ by driving the innovations and gauge processes with appropriate pseudo-random numbers, as is commonly done in quantum optics and atomic physics~\cite{Carm99}. Similarly, if we define $\bar{j}_t(X) = \bar{U}_t^\dag X\bar{U}_t$ using~\eqref{eq approximate qsde} we can construct the `idealized' pure state propagator assuming homodyne measurement of $\bar{Y}_t = \bar{j}_t(A_t+A_t^\dag)$ as
\begin{equation}\label{eq approximate pure2}
 d\bar{v}_t = \Bigg\{\alpha \Pi_{12}d\bar{Y}_t - \frac{|\alpha|^2}{2}Qdt\Bigg\}\bar{v}_t,\qquad \bar{v}_0 \in \mathcal{Q}.
\end{equation}
It is straightforward to show that the idealized filter~\eqref{eq approximate pure2} represents a finite-time unraveling of an ideal $\Pi_{12}$ projective measurement as discussed in~\cite{Stoc04}. We can alternatively think of~\eqref{eq approximate pure2} as a {\it reduced filter} for analyzing the homodyne photocurrent in our two-cavity setup, which exploits adiabatic elimination for a reduction in variable count. In particular we can track the parity of the two qubits (approximately, but with little computational effort) by driving~\eqref{eq approximate pure2} with a given homodyne photocurrent in place of $d\bar{Y}_t$. In an experimental scenario we would use the measured photocurrent; below we will also examine the results of driving~\eqref{eq approximate pure2} with simulated photocurrents generated by the physical Eq.~\eqref{eq pure2}.

To demonstrate that our physical setup indeed realizes an approximate parity measurement, we use~\eqref{eq pure2} to simulate entanglement generation from separable initial states of the atom-cavity systems. We use parameters that should be achievable in a Fabry-Perot cavity/cold Cs atom system with mm-scale dielectric mirrors, $\{g,\kappa^2/2,\gamma^2/2,\alpha\} = \{20,4.5,0.5,0.2\}$. We numerically integrate~\eqref{eq pure2} from the initial state $v_0 = 2^{-1}((|+\rangle+|-\rangle)\otimes|0\rangle)^{\otimes2}$. As can be seen in Fig.~\ref{Fig 2Cavity project}, the system begins with equal projections on the $2^{-1/2}(|+\pm\rangle+|-\mp\rangle)$ joint atomic states and the expected homodyne signal vanishes. The laser is then adiabatically switched on and the conditional state gradually projects into an atomic parity subspace with the corresponding $\langle dY_t^{(0)}\rangle\approx\pm2\alpha$. After a fixed time $\gamma^2t=80$ the probe is adiabatically switched off. While the probe is on, the non-zero overlap with atomic Bell states not present in the initial state is caused partially by entanglement with the cavity mode states, but the residual expectation of such Bell states after the laser is switched off is wholly due to the accumulation of weak measurements performed \emph{within} the atomic parity subspaces by the field modes (see below).

\begin{figure}[tb]
\begin{center}
\includegraphics[width=3in]{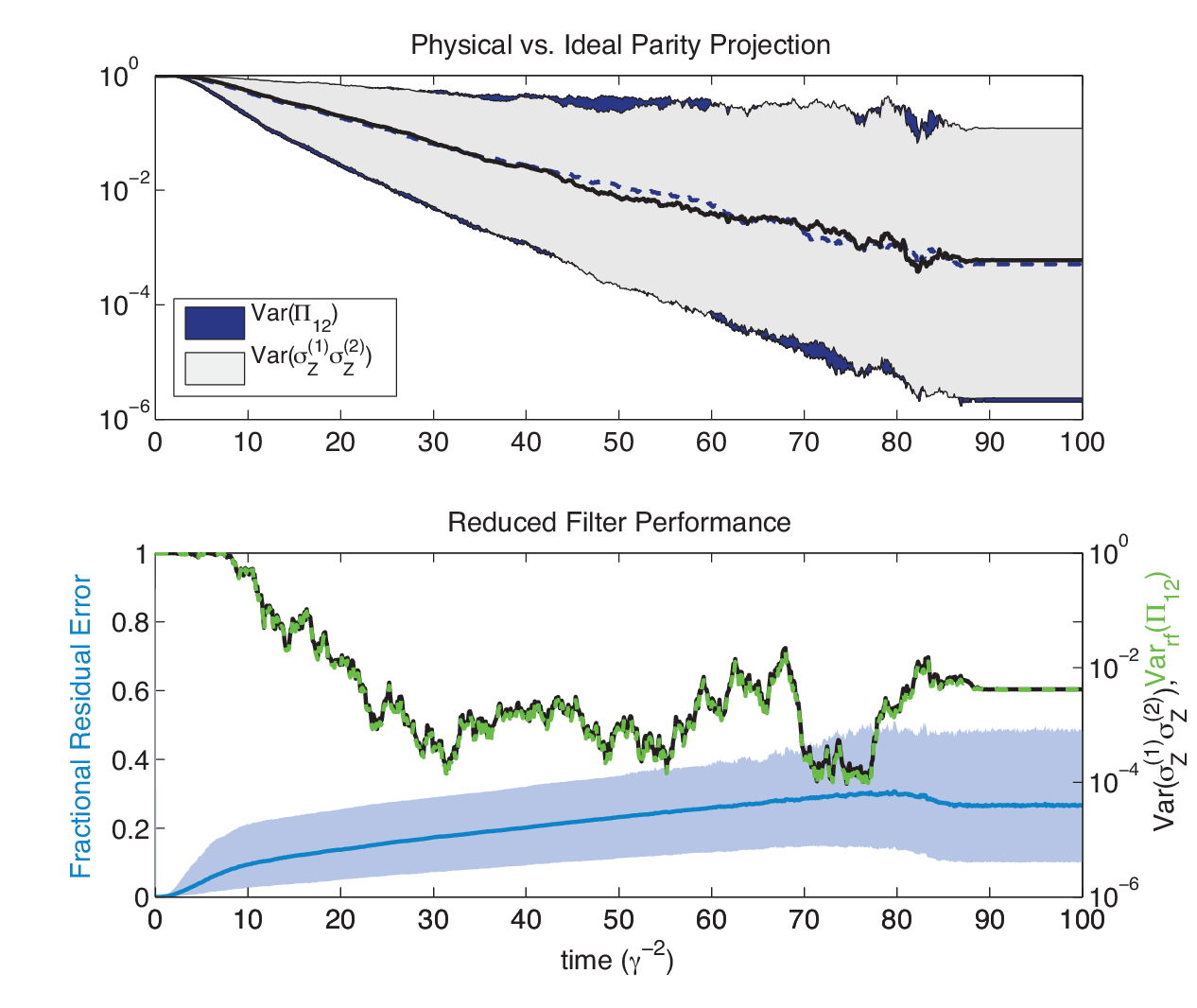}
\vspace{-3mm}
\caption{\label{Fig 2Cavity comparison}
Summary of 1000 Bell-state projection simulations.  The top graph compares the statistics of the ideal and physical parity variance.  The mean of the $\text{Var}(\sigma_Z^{(1)}\sigma_Z^{(2)})$ trajectories is shown as a solid, black line; grey shading indicates the one-standard deviation range of trajectories above and below the mean.  The mean and standard deviation range of $\text{Var}(\Pi_{12})$ are shown by the dotted line and blue shading.  The bottom graph highlights the performance of the reduced filter.  Referenced to the right axis, a representative $\text{Var}(\sigma_Z^{(1)}\sigma_Z^{(2)})$, $\text{Var}_{rf}(\Pi_{12})$ trajectory pair are shown in solid black and dotted green, respectively.  The blue-grey line and shading, referenced to the left axis, depicts the mean and standard deviation range of the fractional residual error over all trajectories.}
\end{center}
\vspace{-0.1in}
\end{figure}

We now compare~\eqref{eq pure2} and~\eqref{eq approximate pure2} by two different computational procedures. First, we compare independent simulations of the idealized projective measurement  represented by Eq.~\eqref{eq approximate pure2}  and the approximate projections represented by Eq.~\eqref{eq pure2}.  To this end, we construct trajectories of the variance of parity operators: $\text{Var}( \sigma_Z^{(1)}\sigma_Z^{(2)})$ from $v_t$ in simulations of Eq.~\eqref{eq pure2} (with $\sigma_Z=|e\rangle\langle e|+|+\rangle\langle+|-|-\rangle\langle-|$ distinguishing cavity-coupled and -uncoupled states), and $\text{Var} (\Pi_{12})$ from $\bar{v}_t$ in independent simulations of Eq.~\eqref{eq approximate pure2}.  Some summary statistics from these simulations are depicted in Fig.~\ref{Fig 2Cavity comparison}. The integrations of the idealized filter are initialized with $\bar{v}_0 = 2^{-1}(|u\rangle+ |d\rangle)^{\otimes2}$.  At $t=0$ we begin with $\text{Var}(\sigma_Z^{(1)}\sigma_Z^{(2)})= \text{Var}(\Pi_{12})=1$, and both variances decrease in time as the systems randomly project into one parity subspace or the other.   Indicative of their similar statistics, the $\text{Var}( \sigma_Z^{(1)}\sigma_Z^{(2)})$ and $\text{Var} (\Pi_{12})$ trajectory ensembles largely overlap at all times.  Moreover, it can be shown that the excited state population remains small at all times, $(\frac{\alpha\kappa}{g})^2$, and that the atomic dynamics are principally constrained to the two ground states. 

Second, we assess how well ~\eqref{eq approximate pure2} performs as a reduced filter for analyzing the physical system.  In this case, we construct trajectories of $\text{Var}_{rf}(\Pi_{12})$ by integrating $\bar{v}_t$ with photocurrents simulated from Eq.~\eqref{eq pure2}.  A representative $\text{Var}_{rf}(\Pi_{12})$,$\text{Var}(\sigma_Z^{(1)}\sigma_Z^{(2)})$ trajectory pair in Fig.~\ref{Fig 2Cavity comparison} suggests the accuracy of the reduced parity estimate.  Underlaying this are the statistics of the fractional residual error from 1000 such pairs: $|\text{Var}(\sigma_Z^{(1)}\sigma_Z^{(2)})-\text{Var}_{rf}(\Pi_{12})|/\text{Var}(\sigma_Z^{(1)}\sigma_Z^{(2)})$.  Although the range of $\text{Var}(\sigma_Z^{(1)}\sigma_Z^{(2)})$ spans 5 orders of magnitude, the reduced filter performs well, tracking this physical parity estimate to within a factor of 2 in every shot.

Now consider $\text{Var}(\sigma_X^{(1)}\sigma_X^{(2)})$ ($\sigma_X= |-\rangle\langle+|+|+\rangle\langle-|$), which serves as a measure of the distinguishability of the atomic states within each parity subspace: $\text{Var}(\sigma_X^{(1)}\sigma_X^{(2)})>0$ indicates an imperfectly prepared atomic Bell state. The idealized description~\eqref{eq approximate pure2} makes no measurement within each parity subspace. $v_t$, however, is capable of (weakly) distinguishing states within parity subspaces through many different mechanisms. A lower bound $\text{Var}(\sigma_X^{(1)}\sigma_X^{(2)})>1-\exp(-8\alpha^2/\kappa^2)$ is set by the non-separability of the atomic and cavity states in equilibrium. As the probe intensity is switched off, the systems factorize again. Spontaneous emission events, although rare (in each time step they occur with probability $\approx(\frac{\gamma\alpha\kappa}{g})^2 dt$), completely destroy any Bell state entanglement. But even if there are no spontaneous emission events the probe alone can make weak measurements within the parity subspaces in steady state. For exactly identical atom-cavities, the even parity subspace has a weak constant decoherence mechanism: the expected homodyne increment $dY_t^{(0)}$ of the $|d\rangle^{\otimes2}$ state is $2\alpha dt$, but is slightly less than that for the $|u\rangle^{\otimes2}$ state because of expected atomic scattering. If the cavities are estimated to be in the state $\cos(\theta)|u\rangle^{\otimes2}+\sin(\theta)|d\rangle^{\otimes2}$, the $\theta$-information content of the next $dY_t^{(0)}$ can be quantified by its Fisher information $\mathcal{I}(\theta) = (2\alpha\sin(2\theta)(\frac{\gamma\kappa}{g})^2)^2dt$. The larger $\mathcal{I}(\theta)$ is, the more likely our estimate of $\theta$ will change because of $dY_t^{(0)}$. Note that $\mathcal{I}(\theta)$ is maximized for the Bell states in this parity subspace, making them especially fragile even in steady state. Of course identical cavity systems are an idealization, but while asymmetries between the cavities can decrease the fragility of the even-parity subspace they increase that of the odd-parity subspace. In general, probe fluctuations cause decoherence as all $|\pm\pm\rangle$ states respond differently to perturbations. Symmetric-cavity considerations of $\text{Var}(\sigma_X^{(1)}\sigma_X^{(2)})$ are thus useful as best- and worst-case scenarios of steady state, probe-induced decoherence.

In summary, the close approximation of ideal parity measurement by a setup based on coherent probe fields and homodyne detection establishes the two cavity system as a continuous-time alternative to the common quantum-circuit building block of sequential controlled-not gates with an ancillary qubit. In this paper we have demonstrated the use of an adiabatic elimination theorem for QSDE's and numerical simulation of derived quantum filtering equations to assess the performance of our proposed implementation of continuous parity measurement in both ideal and realistic parameter regimes.

\begin{acknowledgments} This research is supported by the ARO under W911NF-06-1-0378, by the ONR under N00014-05-1-0420, and by an HP Labs Innovation Research Award.
\end{acknowledgments}


\begin{thebibliography}{99}

\bibitem{Pres98} J.~Preskill, 
    Proc.\ R.\ Soc.\ Lond.\ A {\bf 454}, 385 (1998).

\bibitem{Gott97} D.~Gottesman, 
    arXiv:quant-ph/9705052.

\bibitem{Ahn02} C.~Ahn, A.~C.~Doherty and A.~J.~Landahl, 
    Phys.\ Rev.\ A {\bf 65}, 042301 (2002).

\bibitem{Ahn03} C.~Ahn, H.~M.~Wiseman and G.~J.~Milburn, 
    Phys.\ Rev.\ A {\bf 67}, 052310 (2003).

\bibitem{Ahn04} C.~Ahn, H.~Wiseman and K.~Jacobs, 
    Phys.\ Rev.\ A {\bf 70}, 024302 (2004).

\bibitem{Saro04} M.~Sarovar, C.~Ahn, K.~Jacobs and G.~J.~Milburn, 
    Phys.\ Rev.\ A {\bf 69}, 052324 (2004).

\bibitem{VanH05} R.~van~Handel and H.~Mabuchi, 
    arXiv:quant-ph/0511221v1 (2005).

\bibitem{Ores07} O.~Oreshkov and T.~A.~Brun, 
    Phys.\ Rev.\ A {\bf 76} 022318 (2007).

\bibitem{Chas08} B.~A.~Chase, A.~J.~Landahl and JM~Geremia, 
    Phys.\ Rev.\ A {\bf 77}, 032304 (2008).

\bibitem{Yama06} F.~Yamaguchi, K.~Nemoto and W.~J.~Munro, 
    Phys.\ Rev.\ A {\bf 73}, 060302R (2006).

\bibitem{Duan04} L.~M.~Duan and H.~J.~Kimble, 
    Phys.\ Rev.\ Lett. {\bf 92}, 127902 (2004).

\bibitem{Bout07} L.~Bouten, R.~van~Handel and M.~R.~James, 
    SIAM J.\ Control Optim.\ {\bf 46}, 2199 (2007).

\bibitem{BoSi08} L.~Bouten and A.~Silberfarb, 
    Commun.\ Math.\ Phys.\ {\bf 283}, 491 (2008).

\bibitem{BoVHSi08} L.~Bouten, R.~van~Handel and A.~Silberfarb, 
    J.\ Functional Anal.\ {\bf 254}, 3123 (2008).
    
\bibitem{Carm99} H.~J.~Carmichael, in {\it Quantum Future: From Volta and Como to the Present and Beyond}, proceedings of the 10th Max Born Symposium, Przesieka, Poland, edited by Ph.~Blanchard and A.~Jadczyk (Springer Lecture Notes in Physics, 1999).

\bibitem{Stoc04} J.~K.~Stockton, R.~van~Handel and H.~Mabuchi, 
    Phys.\ Rev.\ A {\bf 70}, 022106 (2004).

\end{thebibliography}
\end{document}